\documentclass[aps,prl,twocolumn,groupedaddress]{revtex4-2}
\usepackage[utf8]{inputenc}
\usepackage{amsmath}
\usepackage{amsfonts}
\usepackage{multirow}
\usepackage{graphicx}
\usepackage{array, booktabs, makecell}
\usepackage{siunitx, mhchem}
\usepackage{float}

\begin{document}


\title{Multiplexed long-range electrohydrodynamic transport and nano-optical trapping with cascaded bowtie photonic crystal nanobeams}


\author{Sen Yang$^{1,2,*}$}
\email[]{These authors contribute equally to this work.}
\author{Joshua A. Allen$^{1,2,*}$}
\author{Chuchuan Hong$^{1,3}$}
\author{Kellen P. Arnold$^{1,2}$}
\author{Sharon M. Weiss$^{1,2,3}$}
\author{Justus C. Ndukaife$^{1,2,3,\dagger}$}
\email[]{justus.ndukaife@vanderbilt.edu}
\affiliation{$^{1}$Vanderbilt Institute of Nanoscale Science and Engineering, Vanderbilt University, Nashville, Tennessee 37235, USA}
\affiliation{$^{2}$Interdisciplinary Materials Science, Vanderbilt University, Nashville, Tennessee 37235, USA}
\affiliation{$^{3}$Department of Electrical and Computer Engineering, Vanderbilt University, Nashville, Tennessee 37235, USA}


\date{\today}

\begin{abstract}
Photonic crystal cavities with bowtie defects that combine ultra-high Q and ultra-low mode volume are theoretically studied for low-power nanoscale optical trapping. By harnessing the localized heating of the water layer near the bowtie region, combined with an applied alternating current electric field, this system provides long-range electrohydrodynamic transport of particles with average velocities of 30 $\mathrm{\mu m/s}$ towards the bowtie region on demand by switching the input wavelength. Once transported to a given bowtie region, synergistic interaction of optical gradient and attractive negative thermophoretic forces stably trap a 10 nm quantum dot in a potential well with a depth of 10 $k_\mathrm{B}T$ using a mW input power. 
\end{abstract}


\maketitle

Nanoscale optofluidic cavities incorporating plasmonic and photonic crystal resonators have recently emerged as a powerful platform for chemical and biological sensing \cite{r49}, nanomanipulation \cite{r11} and optical nano-assembly \cite{r28,r50}. Owning to the nanoscale feature of the cavities, the directed transport of nanoparticles and biomolecules to the region of highest electromagnetic field enhancement is critical to device performance. Though plasmonic nanocavities can support localized electromagnetic hotspots, they suffer from intrinsic material loss that gives rise to low quality factor (Q factor) resonances with broad spectral linewidths. This makes realizing multi-resonant plasmonic cavities for wavelength switchable trapping and long-range particle transport extremely challenging. Dielectric photonic crystal (PhC) cavities, on the other hand confine light by means of a defect in an otherwise periodic arrangement of high index dielectric photonic structures \cite{r13,r19} leading to low-loss, narrow linewidth resonances that can be leveraged for wavelength switchable trapping applications. To date, the deterministic transport of particles for interaction with the enhanced field near resonant PhC cavities has been met with challenges. Prior reported attempts to achieve transport of particles to PhC cavities rely on pressure-driven flow \cite{r15,r21}. Unfortunately, this has limited particle capture rate because only the particles in the fluid boundary layer near the cavity can interact with the electromagnetic hotspots. Additionally, such pressure driven flow does not provide the mechanism for actively transporting particles from one nano-cavity to the next.\\
\indent In this letter, we investigate the physics of light-induced near-field trapping, attractive negative thermophoresis and long-range electrohydrodynamic transport of nanoparticles in PhC cavities for directional delivery of particles and trapping at the cavity region by switching the input wavelength.\par
\begin{figure}[htp!]
\includegraphics[scale=0.325]{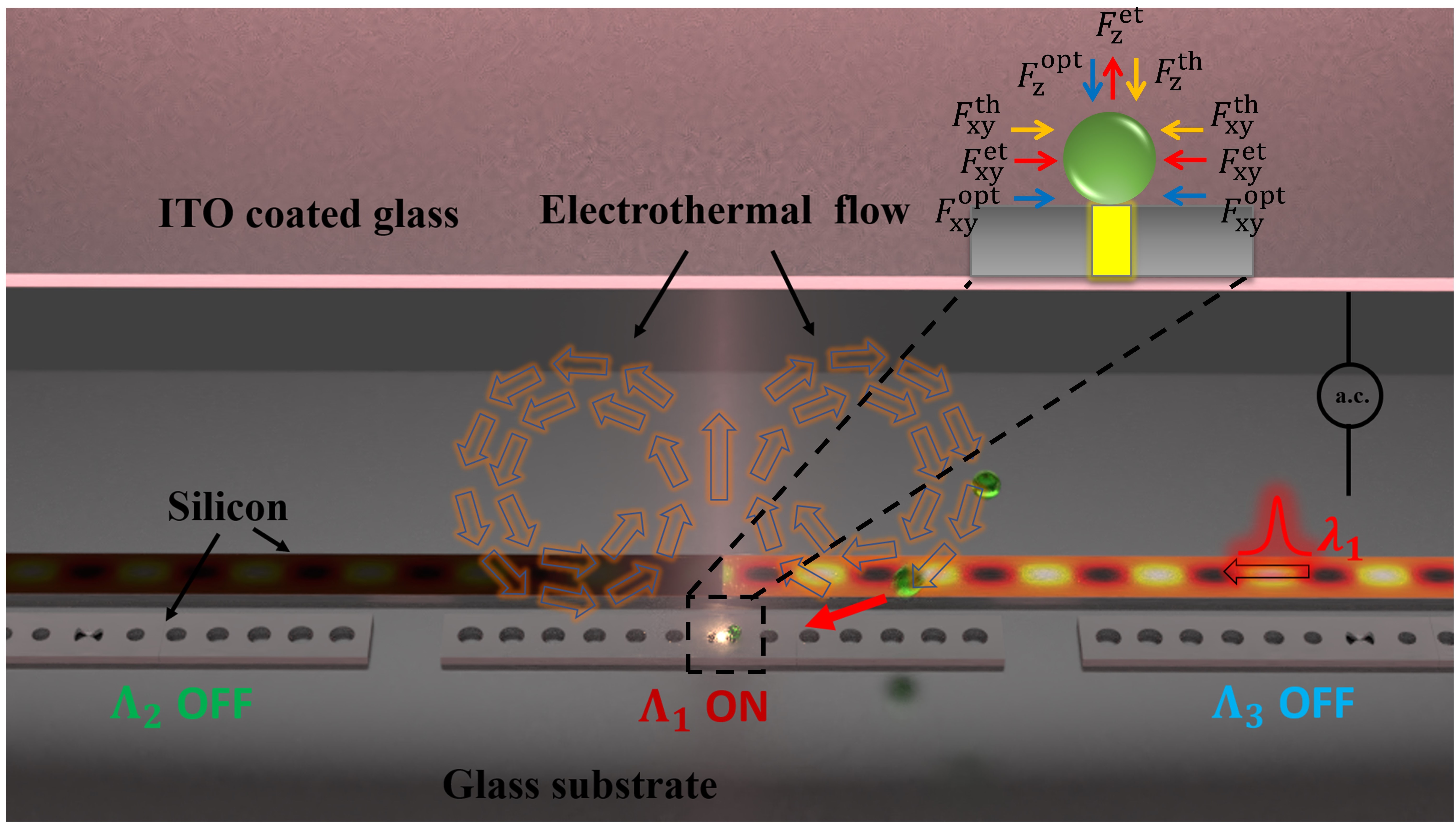}
\caption{\label{fig1} Schematic of the PhC-based multiplexed long-range electrohydrodynamic transport and trapping system. The inset shows the forces experienced by a particle trapped at the bowtie. Here “opt”, “et” and “th” denote “optical force”, “thermophoretic force” and “electrothermal force”, respectively.}
\end{figure}
\begin{figure}[h!]
\includegraphics[scale=0.4]{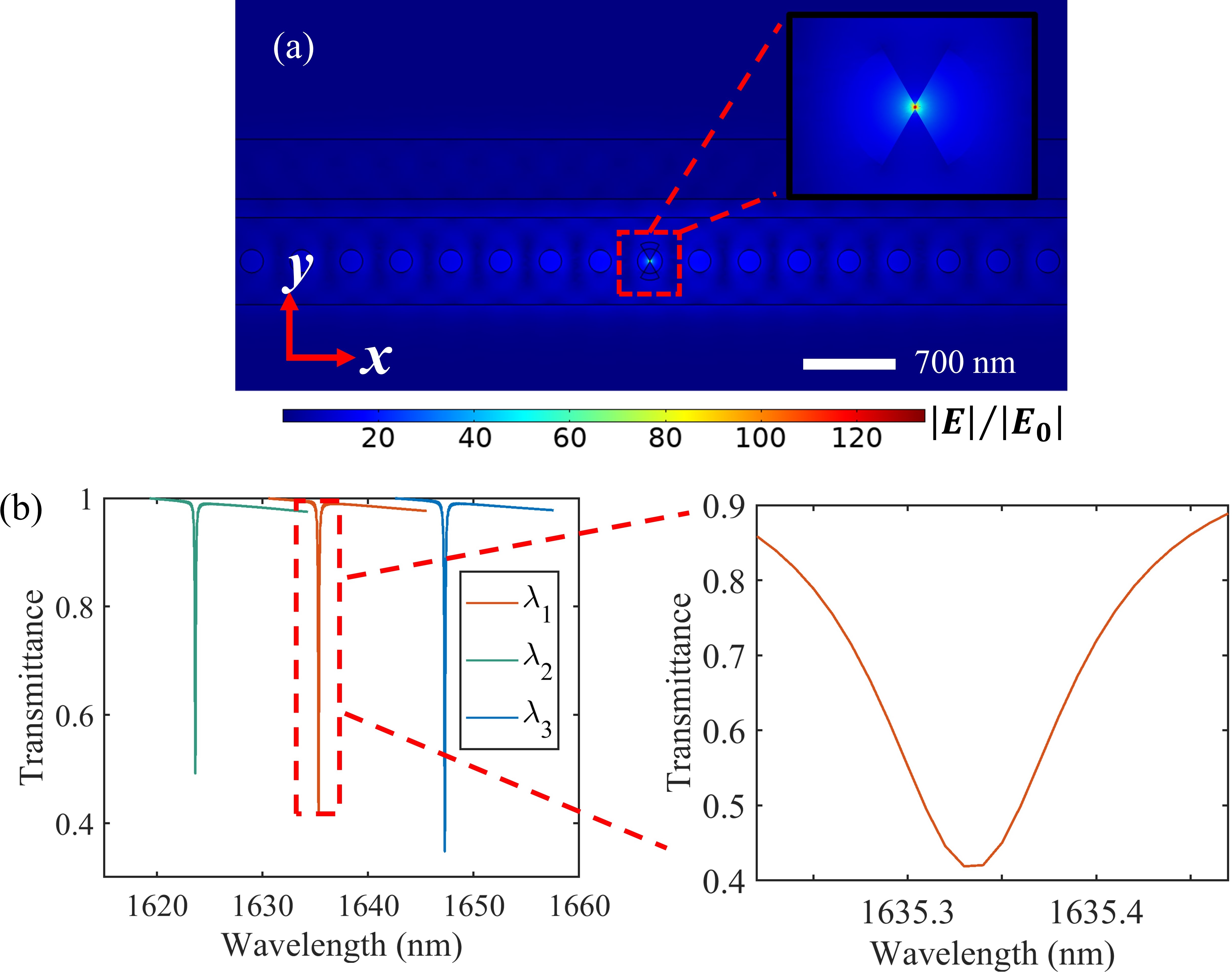}
\caption{\label{fig2} (a) Electric field enhancement distribution of the BPCN cavity on resonance. The field enhancement is calculated by normalization to the amplitude of the electric field of the input fundamental TE mode in the bus waveguide. Inset: Zoom-in view of the bowtie region. (b) Transmission spectra for the three BPCNs, demonstrating the multi-resonant property of the system. The resonant wavelengths are $\lambda_1 = 1635.33  \mathrm{nm}, \lambda_2 = 1623.65 \mathrm{nm}, \lambda_3 = 1647.28 \mathrm{nm}$.}
\end{figure}

As shown in Figure 1, a PhC nanobeam with a bowtie defect at the center is placed beside a bus waveguide to enable evanescent coupling from the side. For multiplexed long-range nanoparticle transport and trapping across cavities, we design the system to contain ($n = 3$) engineered bowtie PhC nanobeams (BPCNs) cascaded along the bus waveguide, each having a specific resonance wavelength. Light is coupled into the device through the bus waveguide and then extremely localized in the bowtie defect. The figure illustrates that a fundamental TE mode with the wavelength of $\lambda_1$ is propagating in the bus waveguide and coupled to the middle BPCN ($\Lambda_1$), inducing the cavity resonance, and switching the cavity to the "ON" state. With a bus waveguide width of 480 nm and coupling gap of 150 nm, the loaded Q of the side-coupled BPCN is $\mathrm{1.5 \times 10^4}$ and the coupling efficiency \cite{r31} is 62$\mathrm{\%}$ in a water environment. The peak electric field amplitude is 138 times higher in the bowtie compared to that of the light input into the bus waveguide, resulting in an electromagnetic field intensity enhancement of $\sim19000$. For details about the design of the BPCNs, please see Supporting Information S1. The electric field profile of the side-coupled BPCN is shown in Figure \ref{fig2}(a), where the extreme light localization at the center of the bowtie can be easily observed. To obtain different resonance wavelengths for the other two BPCNs, we slightly adjust the period of the PhC holes by -4 nm and +4 nm, respectively. It is imperative to note that in practical experiments, it is not necessary to carefully engineer the dimensions of the BPCN to obtain different resonant wavelengths since fabrication uncertainties can naturally introduce resonance shift due to the high Q characteristic of the bowtie PhC. Moreover, resistive heaters can be incorporated to carefully tune the PhC cavities to the desired resonance wavelengths. Figure \ref{fig2}(b) shows the transmission spectra presenting the multi-resonant property of our system. \par
\begin{figure}[h!]
\includegraphics[scale=0.45]{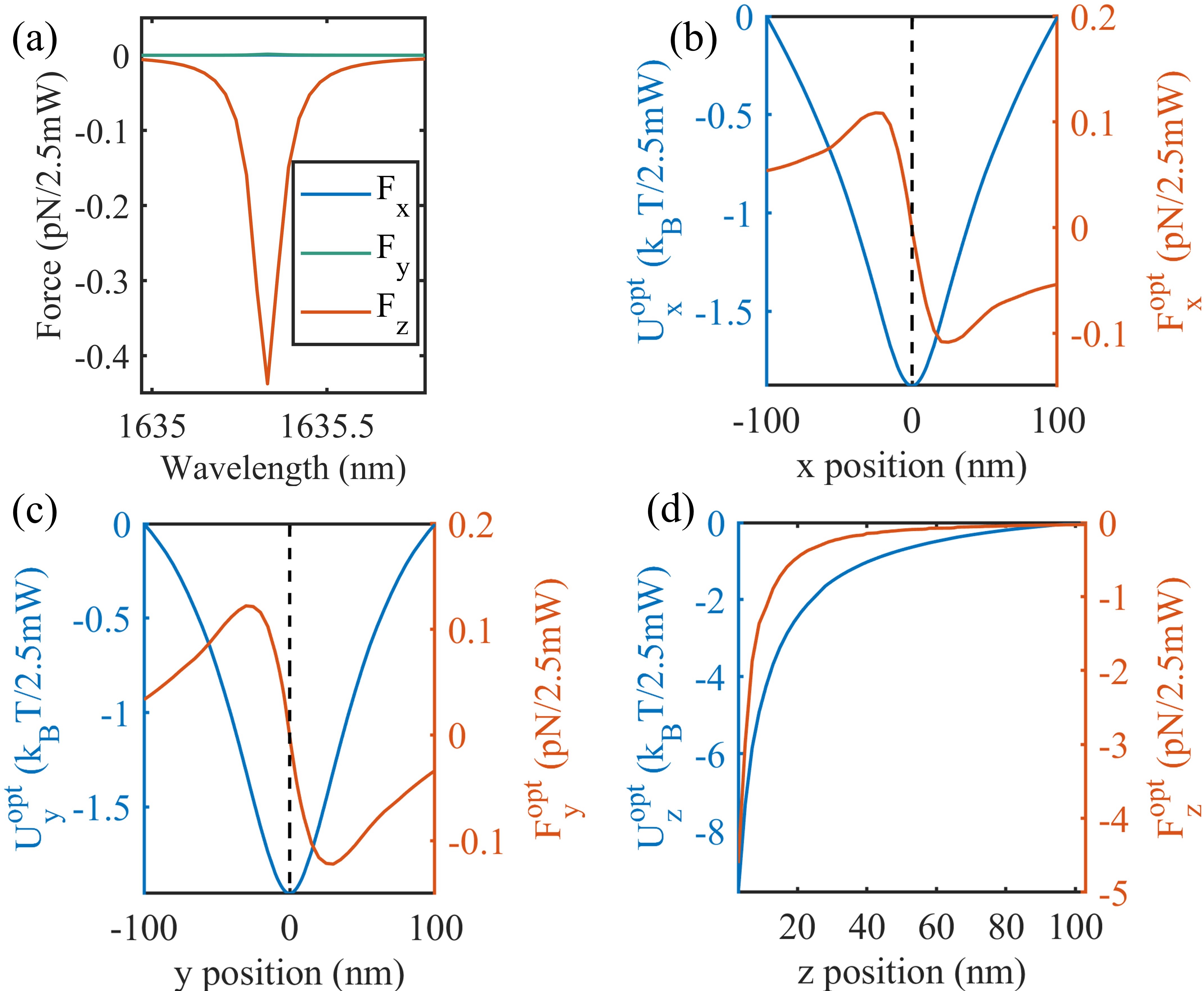}
\caption{\label{fig3} Optical trapping characterization for a 10nm PbSe quantum dot placed 21 nm above the bowtie surface. (a) Trapping force spectra for the quantum dot. (b) – (d) Trapping potential as well as trapping forces when moving the quantum dot along the \emph{x}, \emph{y} and \emph{z} directions, respectively.  The vertical dashed lines in (b) and (c) denote the center of the bowtie. }
\end{figure}

Next, we characterize the optical trapping performance of the optimized BPCN system. An enhanced optical gradient trapping force requires a spatially confined electromagnetic field, which is provided by the bowtie. The time averaged optical force exerted on a nanoscale object is calculated by integrating the Maxwell’s stress tensor (MST) \cite{r32} over an arbitrary surface enclosing the nanoscale object, which is given by
\begin{eqnarray}\label{eq1}
\langle\mathbf{F}\rangle=\oint_{S}\langle\overrightarrow{\mathbf{T}}\rangle \cdot d \mathbf{S},
\end{eqnarray}
where $\langle\overrightarrow{\mathbf{T}}\rangle$ is the time averaged Maxwell’s stress tensor given by:
\begin{eqnarray}\label{eq2}
\langle\overrightarrow{\mathbf{T}}\rangle=\frac{1}{2}\operatorname{Re}\left[\varepsilon \mathbf{E} \mathbf{E}^{*}+\mu \mathbf{HH}^{*}-\frac{1}{2}\left(\varepsilon|\mathbf{E}|^{2}+\mu|\mathbf{H}|^{2}\right)I\right]
\end{eqnarray}
Here $\mathbf{EE}^{*}$ and $\mathbf{HH}^{*}$ are the outer products of the fields; \emph{I} is the identity matrix; and $\varepsilon$ and $\mu$ are the permittivity and the permeability of the medium surrounding the object, respectively. The effective transverse trapping potential resulting from the optical force is given by \cite{r33}
\begin{eqnarray}\label{eq3}
U\left(\mathbf{r}_{0}\right)=\int_{\infty}^{r_0} \mathbf{F}(\mathbf{r}) d \mathbf{r}.
\end{eqnarray}
Figure \ref{fig3}(a) illustrates the force spectra of a 10 nm diameter ($D$ = 10 nm) PbSe quantum dot (refractive index \emph{n} = 4.73 + 0.24i \cite{r35}) positioned 21 nm above the center of the bowtie, showing a strong pulling force along the z direction (0.44 pN/2.5mW). This trapping force is at least an order of magnitude higher than those achieved using Mie-resonant dielectric nanoantenna \cite{r34} and dielectric nanoantenna supporting anapole states \cite{r46}. Figure \ref{fig3}(b) to \ref{fig3}(d) show the trapping potentials as well as the corresponding trapping forces when moving the quantum dot along the \emph{x}, \emph{y} and \emph{z} directions, respectively. The depth of the trapping potential well is around 2 $k_\mathrm{B}T$/2.5mW for the $x$ and $y$ directions and 10 $k_\mathrm{B}T$/2.5mW for the \emph{z} direction. Here $k_\mathrm{B}$ is the Boltzmann constant. This provides sufficient potential depth to stably confine the nanoparticle near the bowtie region.\par
\begin{figure}[h!]
\includegraphics[scale=0.6]{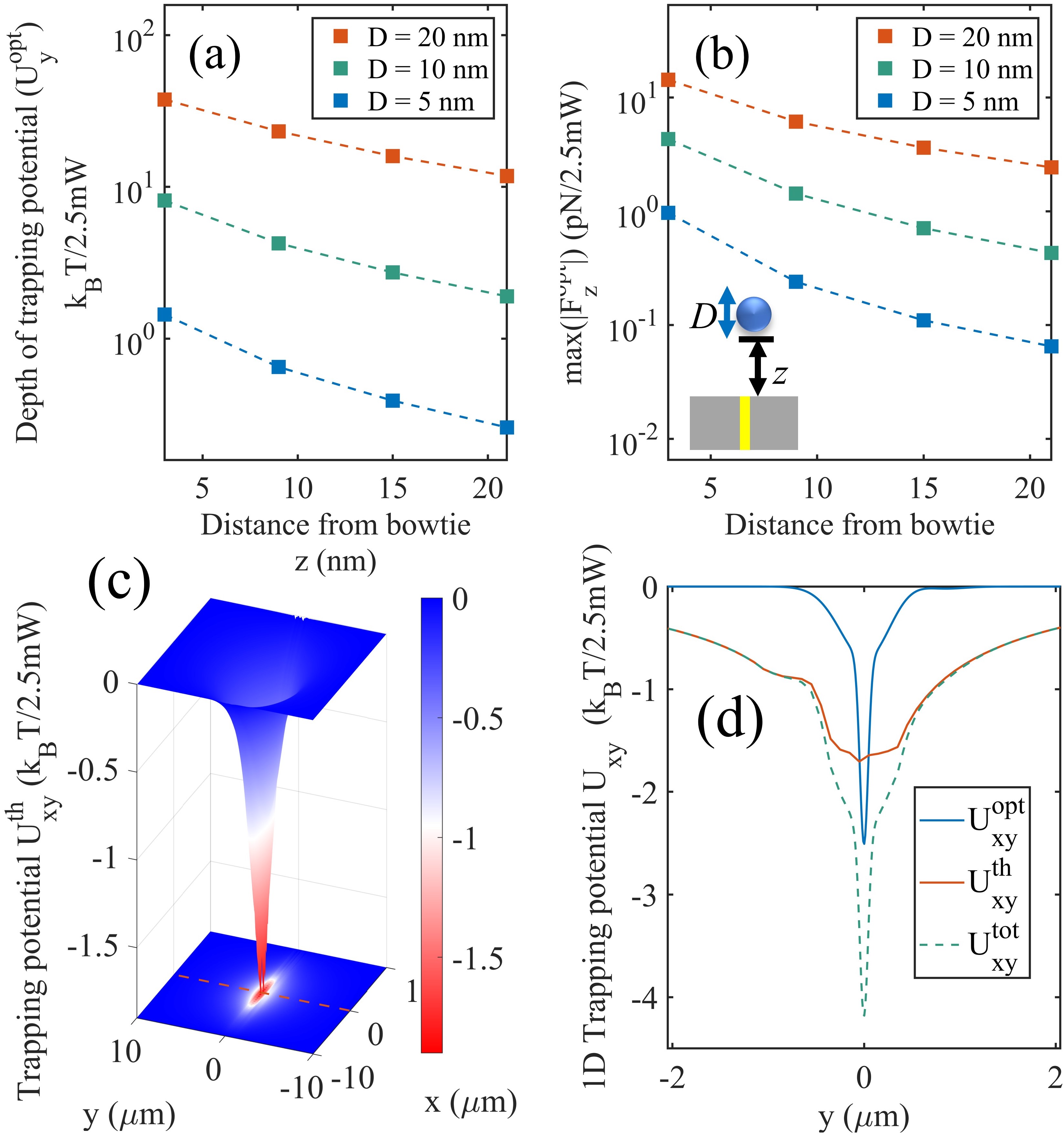}
\caption{\label{fig4} Depth of the trapping potential well along the y direction ($U_{y}^\mathrm{opt}$) due to $F_{y}^\mathrm{opt}$ (a) and the maximum absolute value of the pulling force $F_{z}^\mathrm{opt}$ (b) at different distances (z) from the bowtie surface (shown in the inset of (b)). z is varied as 3, 9, 15, and 21 nm. Both y axes are in log scale. (c) Transverse thermophoretic trapping potential for 10 nm particle at \emph{z} = 21 nm. The red dash line denotes the trace in (d). (d) Transverse trapping potential for the optical trapping and thermophoretic trapping potentials along the y axis.}
\end{figure}

Next, we explore the impact of the particle size and the vertical distance between the particle and the bowtie on the optical forces. We consider three different particle diameters (\emph{D} = 5, 10, and 20 nm) for the quantum dot. Figure \ref{fig4}(a) shows the depth of the optical trapping potential well along the $y$ direction ($U_{y}^\mathrm{opt}$) with respect to $F_{y}^\mathrm{opt}$ and Figure \ref{fig4}(b) shows the maximum absolute value of the pulling force $F_{z}^\mathrm{opt}$ as a function of the distance (\emph{z}) measured from the quantum dot bottom to the bowtie surface (shown in the inset), respectively.
Given the stable trapping requirement of 10 $k_\mathrm{B}T$ for the trapping potential well, Figure \ref{fig4}(a) indicates that the minimum power required for stable trapping of the three particles considered along the y direction is around 17 mW, 3 mW and 0.7 mW, respectively, when $z$ is 3 nm, which is less than a half of the power required by a plasmonic nanoaperture \cite{r33} taking into account the refractive index differences. Figure \ref{fig4}(b) shows that the maximum pulling force along the \emph{z} direction is 0.97 pN/2.5mW, 4.31 pN/2.5mW and 14.30 pN/2.5mW for the three particles considered at the same distance (\emph{z} = 3 nm). This pulling force drops exponentially as the particle moves away from the bowtie. Additional discussions on optical forces can be found in Supporting Information S3.\\
\indent Optical power dissipated in the water layer near the bowtie establishes a thermal gradient. To calculate the temperature field distribution, the computed electric field distribution is used to determine the heat source density, which gives the heat dissipated per unit volume and is expressed as \cite{r43,r56}
\begin{eqnarray}\label{eq4}
q(\mathbf{r})=\frac{1}{2} \operatorname{Re}\left[\mathbf{J}_{d}^{*}(\mathbf{r}) \cdot \mathbf{E}(\mathbf{r})\right]=\frac{\omega}{2} \operatorname{Im}(\varepsilon)|\mathbf{E}(\mathbf{r})|^{2},
\end{eqnarray}
where $\mathbf{J}_{d}=i\omega\mathbf{D}$ with $\mathbf{D}=\varepsilon\mathbf{E}$ is the displacement current density and $\varepsilon$ is the relative permittivity of the specific material. Since water is the only lossy material in this system, the power density dissipated into water serves as the source term in the heat diffusion equation for computation of the temperature around the PhC as well as in the surrounding fluid and substrate. The temperature field in the system is determined by solving the steady-state heat equation given by  
\begin{eqnarray}\label{eq5}
\nabla \cdot\left[-\kappa \nabla T(\mathbf{r})+\rho c_{p} T(\mathbf{r}) \mathbf{u}(\mathbf{r})\right]=q(\mathbf{r}).
\end{eqnarray}
The first term on the left is the heat conduction term, while the second term is the convection term, which depends on the velocity of the fluid. $T(\mathbf{r})$ and $\mathbf{u}(\mathbf{r})$ are spatial temperature and fluid velocity field, respectively, and the material properties $\kappa$, $\rho$ and $c_p$ are thermal conductivity, density and specific heat capacity, respectively.\\
\indent Thermophoresis phenomenon \cite{r54,r55} is the motion of particles or molecules in the presence of thermal gradients and provides an attractive trapping force under negative thermophoresis \cite{r38,r39,r41}. Details of the calculation of the negative thermophoretic force are provided in the Supplementary Information S4. To compare with the optical trapping, Figure \ref{fig4}(c) illustrates the simulated transverse thermophoretic trapping potential ($\mathrm{U}_{xy}^\mathrm{th}$) for 10 nm particle at \emph{z} = 21 nm. The thermophoretic trapping potential has the same order of magnitude as the optical trapping potential shown in Figure 3, whereas the thermophoretic trapping potential well is much broader than the optical one. For emphasis, the transverse optical trapping and thermophoretic trapping potential along the $y$ axis are shown in Figure \ref{fig4}(d). The asymmetry of the thermophoretic trapping potential curve results from the bus waveguide which affects the heat diffusion. The superposition of the optical trapping and the thermophoretic trapping achieves a long range and 1.7 times deeper trapping potential well.\\
\begin{figure}[h!]
\includegraphics[angle=270,scale=0.475]{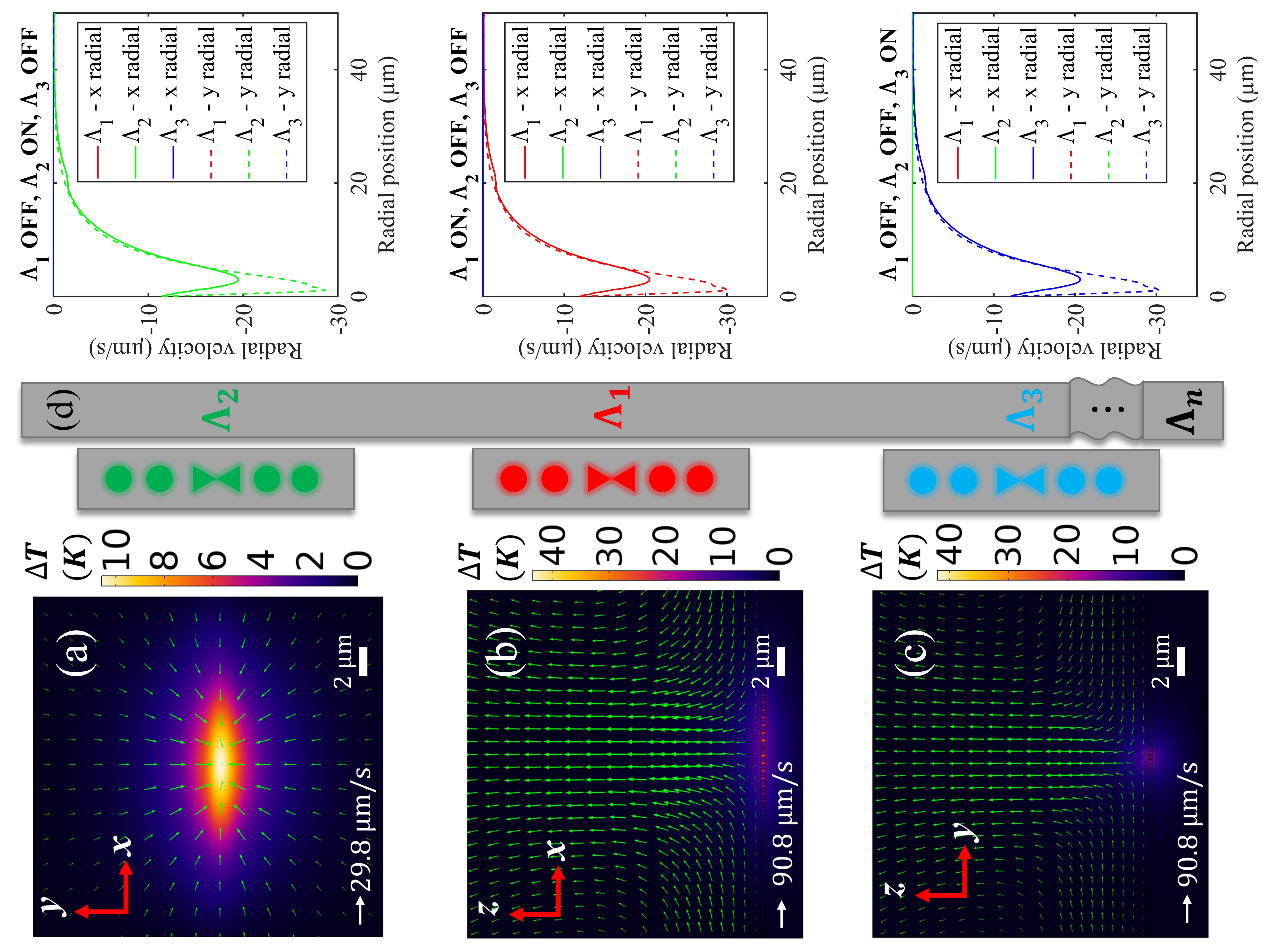}
\caption{\label{fig5} (a) Temperature field distribution of the $xy$ plane 300 nm above the bowtie. The radial velocity vector plot of electrothermal flow induced around the resonant BPCN is superimposed on the temperature profile. Arrow length represents the magnitude of the flow velocity. (b) Temperature field distribution of the $xz$ plane and (c) the $yz$ plane. (d) Illustration of three cascaded BPCNs placed beside a bus waveguide. The right three panels show corresponding radial velocity profiles of the induced electrothermal flow along $x$ and $y$ direction (300 nm above the surface) around the three BPCNs in different states. The electrothermal flow shows a long-range characteristic ($\sim 50 \mathrm{\mu m}$). (a) to (c) corresponds to the $\Lambda_1$ panel.}\par
\end{figure}
Next, we demonstrate the long-range and rapid transport of individual nanoparticles to the vicinity of the bowtie for optical trapping by generating wavelength-dependent electrohydrodynamic microfluidic vortices based on the electrothermal flow effect in a microfluidic channel. The thermal gradient induced in the water layer near the bowtie results in a gradient in the permittivity and electrical conductivity of the water medium near the bowtie. An applied a.c. electric field acts on these gradients to create a volumetric body force in the fluid due to the electrothermal effect \cite{r22,r52}.\\
\indent By leveraging the configuration of the cascaded BPCNs, the body force of the electrothermal flow in our system is not only space dependent, but also \emph{wavelength dependent}. That means the spatial distribution of the local temperature gradient can additionally be controlled by the wavelength of the input light, as shown in Figure \ref{fig5}(d). Following a perturbative expansion \cite{r22}, the wavelength dependent time-averaged electrothermal body force per unit volume at a.c. frequency $\omega$ can be expressed as:
\begin{eqnarray}\label{eq6}
\left\langle F_{\text{ET}}\right\rangle \hat{z}=\frac{1}{2} \varepsilon E_{z}^{2}\left[\frac{\sigma^{2} \varepsilon(\alpha-\gamma)}{\sigma^{2}+\omega^{2} \varepsilon^{2}}-\frac{1}{2} \alpha\right] \frac{\partial T(z, \lambda)}{\partial z} \hat{z},
\end{eqnarray}
\begin{eqnarray}\label{eq7}
\left\langle F_{\text{ET}}\right\rangle \hat{r}=-\frac{1}{4} \varepsilon \alpha E_{z}^{2} \frac{\partial T(r, \lambda)}{\partial r} \hat{r},
\end{eqnarray}
where $\lambda=\lambda_1,\lambda_2,\lambda_3…\lambda_n$ and $n$ is the number of the cascaded BPCNs; $\varepsilon$, $\tau=\varepsilon/\sigma$, $\sigma$ and $\omega$ are the fluid permittivity, charge relaxation time, electrical conductivity, and applied a.c. frequency, respectively; $\alpha$ and $\gamma$ are expressed as $\alpha=(1 / \varepsilon)(\partial \varepsilon / \partial T)$, $\gamma=(1 / \sigma)(\partial \sigma / \partial T)$ and are given as $-0.004 \mathrm{~K}^{-1}$ and $0.02 \mathrm{~K}^{-1}$, respectively \cite{r53}. Eq. \ref{eq6} and Eq. \ref{eq7} describe the axial and radial components of the electrothermal body force.\\
\indent The velocity field distribution of the fluidic flow when a given BPCN is excited is determined from the solution of the incompressible Navier–Stokes equations given by
\begin{eqnarray}\label{eq8}
\rho_{0}(\mathbf{u}(\mathbf{r}) \cdot \nabla) \mathbf{u}(\mathbf{r})+\nabla p(\mathbf{r})-\eta \nabla^{2} \mathbf{u}(\mathbf{r})=\mathbf{F},
\end{eqnarray}
\begin{eqnarray}\label{eq9}
\nabla \cdot \mathbf{u}=0.
\end{eqnarray}
The forcing term $\mathbf{F}$ in Eq. 8 describes the body force per unit volume acting on the fluid element, which is given by Eq. \ref{eq6} and Eq. \ref{eq7}. We note that contribution from buoyancy-driven convection is negligible in this system (see Supporting Information S7 for details).\\
\indent Figure \ref{fig5}(a) shows the temperature field distribution in the \emph{xy} plane 300 nm above the PhC surface. A temperature rise of 10.6 K is observed and the superimposed radial velocity vector plot of the flow shows that the induced electrothermal flow is directed radially inwards towards the thermal hotspot generated by water around the bowtie and serves as a powerful means to deliver suspended particles to the bowtie region. The maximum flow velocity is about 29.8 $\mathrm{\mu m/s}$, which is directed to the bowtie and hence much more efficient than traditional particle delivery methods such as pressure-driven flow and slow Brownian motion. Furthermore, this flow velocity is at least 20 times greater than the 1 $\mathrm{\mu m/s}$ thermoplasmonic convection flow velocity achievable with an array of optimized plasmonic bowtie nanoantenna \cite{r51}. Figure \ref{fig5}(b) and (c) show the temperature profile in the \emph{xz} and the \emph{yz} plane. The maximum temperature rise is 43.6 K under only 2.5 mW input power due to the high field enhancement. The superimposed velocity vectors verify the induced electrothermal vortex flow shown in Figure \ref{fig1}.\\
\indent Figure \ref{fig5}(d) demonstrates the concept of multiplexed nanoparticle transport and nano-optical trapping with the cascaded BPCNs shown in Figure \ref{fig1}. When the input wavelength is tuned to $\lambda_1$, only the middle BPCN ($\Lambda_1$) with a resonance wavelength of $\lambda_1$ is excited while the other two are off resonance and not excited. Therefore, the electrothermal flow is only induced around $\Lambda_1$ (see the radial velocity plot in the middle panel). By integrating over the radial velocity curve when $\Lambda_2$ is on, we estimate that it takes about 8 seconds for the flow to transport a nanoparticle 25 $\mathrm{\mu m}$ from the vicinity of $\Lambda_1$ to $\Lambda_2$. The high Q characteristic of the bowtie PhC permits to integrate multiple BPCNs along the low-loss bus waveguide to provide the means to achieve long range particle hand-off from tens of microns to millimeter scale distances by simply switching the wavelength of the input light. We note that the electrothermal flow along the $y$ direction shows a larger magnitude of velocity in comparison to the electrothermal flow along the $x$ direction.  This is attributed to the asymmetric spatial distribution of the in-plane thermal hotspots. It is evident that there is a higher temperature gradient along the y direction and hence a stronger electrothermal flow velocity in comparison to that along the $x$ direction. The slightly different radial velocity values presented in the three panels are attributed to the different field enhancements of the three BPCNs.\\
\indent We have proposed and systematically studied a cascaded bowtie photonic crystal nanobeam system that can achieve multiplexed long-range electrohydrodynamic transport and optical trapping of nanoscale particles. Compared with traditional 1D photonic crystal nanobeams, our bowtie photonic crystal cavity can more strongly confine and enhance the electromagnetic field while maintaining a high quality factor. The extremely localized field provides a strong field gradient that is ideal for trapping sub-20 nm particles. Furthermore, the localized water absorption near the cavities serves as heat sources to generate negative thermophoresis that can assist in the optical trapping process. Finally, we harness the localized water absorption to induce on-demand electrothermal flow that can efficiently transport nanoparticles to the vicinity of the localized field of the bowtie photonic crystal cavity region for enhanced optical trapping. Our proposed multiplexed platform could enable millimeter scale transport and hand-off of particles across cavities in miniaturized optofluidic chips by simply switching the wavelength. We envision that our system will be a promising platform in many fields of biology and quantum information, such as in single molecule characterization and assembly of single photon sources.\\

\begin{acknowledgments}
J.C.N and S.M.W conceived and guided the project. J.A.A designed the photonic crystal cavity. J.A.A and S.Y performed the wave-optics simulations in Lumerical. S.Y. implemented the photonic crystal design in COMSOL and performed the wave-optics and Multiphysics simulations in COMSOL. C.H contributed to the Multiphysics simulations and K.P.A contributed to the Lumerical simulations. J.A.A wrote the bowtie PhC design section and S.Y. wrote the rest of the manuscript. J.C.N and S.M.W contributed to editing the manuscript.\\

The authors acknowledge financial support from the National Science Foundation (NSF ECCS-1933109) and Vanderbilt University. 
\end{acknowledgments}

\bibliography{references.bib}

\end{document}